\documentclass[journal,comsoc,hidelinks]{IEEEtran}

\usepackage[T1]{fontenc} 
\usepackage{color}
\usepackage{multirow}
\usepackage{array}
\usepackage{amsmath}
\usepackage{amssymb}
\usepackage{amsthm}
\usepackage{mathrsfs}
\usepackage{url}
\usepackage{soul} 
\usepackage{graphicx} 
\usepackage{stfloats}
\usepackage{enumitem}
\usepackage{algorithmic}
\usepackage{algorithm}
\usepackage{overpic}
\usepackage{cite} 
 
\makeatletter
\newcommand{\removelatexerror}{\let\@latex@error\@gobble}
\makeatother

\graphicspath{} 

\def\ie{{i.e.}}
\def\etc{{etc}}

\def\Figure{{Fig. }}

\begin{document}
	\title{Rethinking Modern Communication from Semantic Coding to Semantic Communication}
	\author{Kun~Lu,~Qingyang~Zhou,~Rongpeng~Li,~Zhifeng~Zhao,~Xianfu~Chen,~Jianjun~Wu,~and~Honggang~Zhang
		\thanks{This article is accepted by IEEE Wireless Communications (DOI: 10.1109/MWC.013.2100642).
				Kun Lu, Qingyang Zhou, Rongpeng Li, and Honggang Zhang are with Zhejiang University; Zhifeng Zhao is with Zhejiang Lab, and also
			with Zhejiang University; Xianfu Chen is with VTT; Jianjun Wu is with Huawei Technologies.}
	}

	\maketitle

	\begin{abstract}

	Modern communications are usually designed to pursue a higher bit-level precision and fewer bits while transmitting a message. This article rethinks these two major features and introduces the concept and advantage of semantics that characterizes a new kind of semantics-aware communication framework, incorporating both the semantic encoding and the semantic communication problem. After analyzing the underlying defects of existing semantics-aware techniques, we establish a confidence-based distillation mechanism for the joint semantics-noise coding (JSNC) problem and a reinforcement learning (RL)-powered semantic communication paradigm that endows a system the ability to convey the semantics instead of pursuing the bit level accuracy. 
	On top of these technical contributions, this work provides a new insight to understand how the semantics are processed and represented in a semantics-aware coding and communication system, and verifies the significant benefits of doing so. Targeted on the next generation's semantics-aware communication, some critical concerns and open challenges such as the information overhead, semantic security and implementation cost are also discussed and envisioned.

	\end{abstract}
	
	
	\IEEEpeerreviewmaketitle
	
	\section{Introduction}
	\IEEEPARstart{T}{he} core of communication is to send and receive useful messages. Till now, the most convincing metric used to evaluate a communication system is bit error rate, which encourages an exact recurrence of what has been sent from the transmitter. 
	Reliable though this long-established practice is, it can result in two potential issues - the encoded bit flow may get wasted on the less important or redundant information; and the decoded message may not express the true and important meaning of what is expected to send. The information waste and tremendous overheads in our communication systems, as we can foresee in the following decades, inspire us to rethink and reshape the way we view our communication system, in a semantics-aware manner.

	As our primary goal, the idea of meaningfully transmitting messages was discussed as early in 1949 by Weaver, who extended Shannon’s theory to two extra levels: the semantic level and effective level \cite{shannon1949mathematical}. A semantic communication system is targeted to transmit the semantic information, while an effective one further pursues an efficient and goal-oriented system design. After Weaver, some early researches on semantic information were thereafter proposed \cite{carnap1952outline}. Unfortunately, the idea of communicating semantics did not gain much attention at that time, due to the urgent need of Shannon’s high-rate reliable communication and the lack of opportunity and computing resources to unfold the next level. However at the time when high-rate reliable communication is no longer a critical barrier to communicate, as in beyond 5G (B5G) and targeted 6G era, a semantics-aware and efficient communication scheme is now gaining much more interest.

	In recent few decades, researches on semantics-aware communication are first advanced by progresses on semantic information theory \cite{bao2011towards, juba2011universal}. Some key features adopted in subsequent works, such as shared common knowledge, goal-oriented schemes, and efficiency issues were declared. These innovations, along with recent theoretical conceptions \cite{strinati20216g, shi2021new}, gradually sketch the portrait of modern semantics-aware communication structure. The semantic abstraction nature also suggests that exploring the underlying semantics is promising to provide a new solution to efficient communication, which is usually designed from the service and energy-utility perspective \cite{buzzi2016survey, papavassiliou2020paradigm}, and become the critical cornerstone for next level's goal-oriented communication.
	
\begin{figure*}[t]
		\centering
		\setlength{\abovecaptionskip}{0.cm}
		\includegraphics[width=18.1cm]{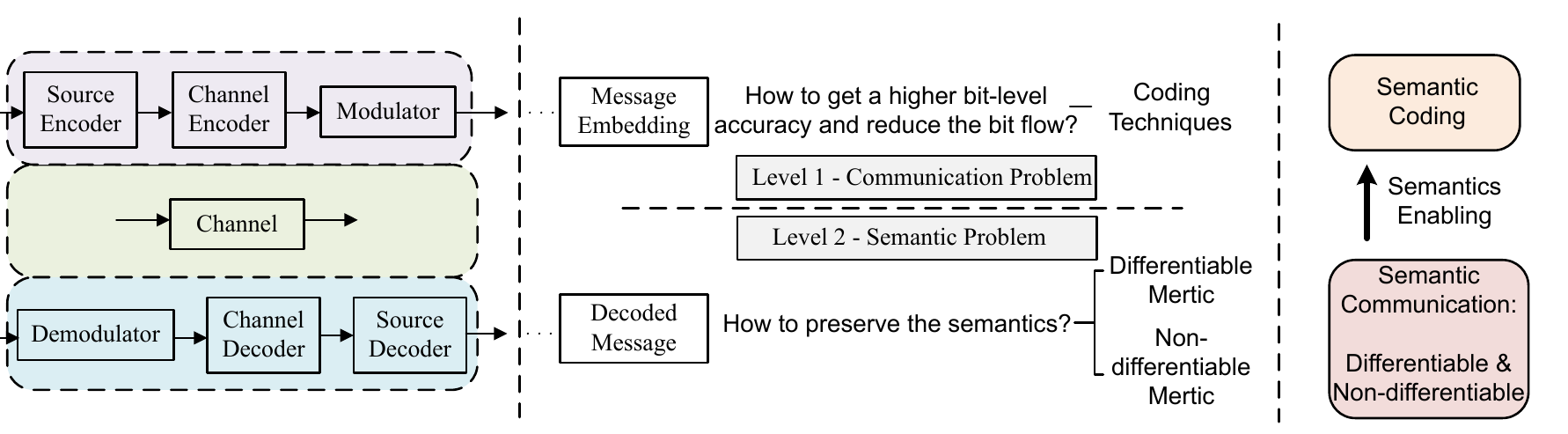}
		\caption{An overview of the joint source and channel coding scheme, and the semantics-aware communication tasks. Left part: basic joint source and channel coding (JSCC) schemes. Middle part: illustration of the targeted communication and semantic problem. Right part: the corresponding solutions.}
		\label{Fig:FIG1}
		\vspace{-0.5cm}
	\end{figure*}

	Nevertheless, practical semantics-aware communication systems remained less-explored until the last 4 to 5 years when machine learning-based calculation, typically the joint source and channel coding schemes (see the left part in \Figure \ref{Fig:FIG1}) became the experimental trend \cite{o2017introduction, bourtsoulatze2019deep}. Among these works, recent research interests in semantics-aware communications can be roughly divided into two branches: semantic coding and semantic communication. 
	The former category concentrates on a reliable and efficient transmission by reducing the length of bits flow or to secure a higher bit accuracy rate \cite{farsad2018deep, xie2021deep}, where the latter is targeted to transmit semantic meanings (typically not designed for a higher bit-level accuracy but for a semantics or goal-execution purpose) \cite{tung2021joint}, as illustrated in the right part of \Figure \ref{Fig:FIG1}. 
	Here, it is noteworthy to point out that semantic coding and semantic communication are not the same concept, although they are often treated the same in the literature. 
	Besides semantic coding to pursue a simple recurrence on the receiver side, a semantic communication system still requires appropriate semantic metrics that guide it toward a semantic level expression (i.e., a redesign on the learning objective). 
	
	Although existing works provide plausible solutions for semantics-aware communication, the drawbacks of these existing methods are also evident: 1) for the communication purpose, the key component of wireless communication - the varying channel state is significantly simplified into a fixed-level AWGN one and not considered carefully (we call this \textit{communication} problem), and 2) for the semantics purpose, existing works either rely on NLP (Natural Language Processing) models to obtain a semantics-aware representation ability instead of carefully rethinking the final goal of semantics transmission, or are too preliminary for a large-scale generalization \cite{tung2021joint} (we call this \textit{semantic} problem). 
	This article aims to provide a holistic view on the semantics-based communication methods and present our new solutions to these interesting yet to be solved problems. 

	As to the first problem, we introduce in this article a joint semantics-noise coding (JSNC) mechanism, so as to automatically adjust the depth of semantic representations according to the varying channel state and varying sentences with different semantic structures. As to the second problem, we further put
	forward a semantic communication system that directly learns from the semantic similarity instead of the semantics-blind cross entropy (CE) loss or mean squared error (MSE), where the semantic similarity could be any task-specified and even non-differentiable one.

	\vspace{-0.3cm}
	\section{Semantic Coding: A Confidence-based Joint Semantics-Noise Coding (JSNC) Manner}

	\subsection{Overview of the Proposed JSNC Method}
\begin{figure}[b!]
		\centering
		\setlength{\abovecaptionskip}{0.cm}
		\includegraphics[width=8.8cm]{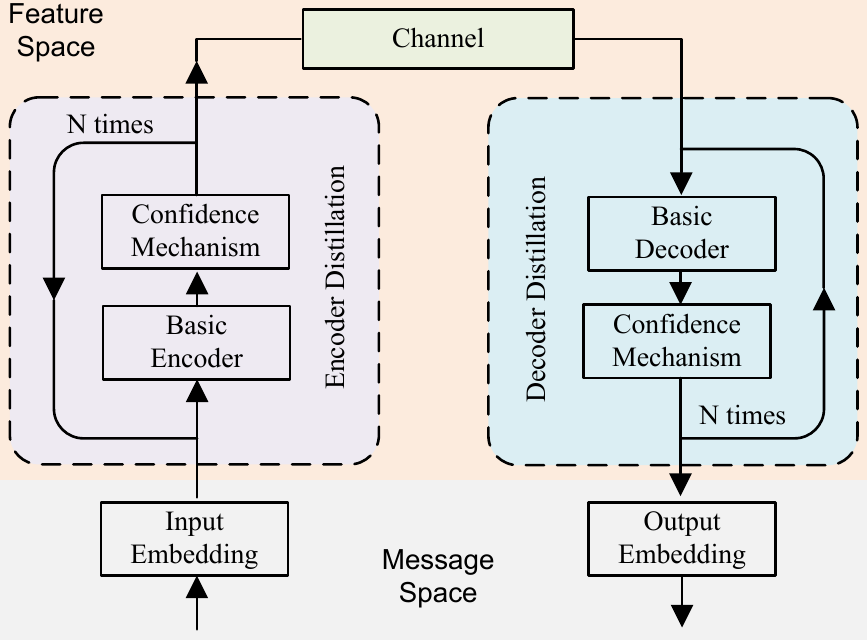}
		\caption{Architecture of the proposed confidence-based joint semantics-noise coding (JSNC) mechanism. The gray part shows the message space, while the orange part represents the feature space. JSNC is adopted in both the encoder (purple block) and decoder (blue block). The message embedding is distilled through the encoder and decoder multiple times, which is controlled by the corresponding learnable confidence module (neural networks).}
		\label{Fig:JSNC}
	\end{figure}

	We provide an overview of the proposed JSNC system in \Figure \ref{Fig:JSNC}. JSNC is significantly different from existing semantic coding ones in two ways: 1) it is designed to address the varying channel situations, which is usually ignored or simplified previously, and 2) JSNC develops an interpretable mechanism that allows to understand why the proposed solution is semantics-aware, and how it interprets the semantic meanings. 
	To achieve the first goal, we propose a distillation mechanism that refines the embeddings in the encoder and decoder, which endows the transceiver a proper semantics-extraction capability to adjust to different sentences and channel situations. As to the second goal, a confidence-based mechanism that evaluates the quality of semantic representation and guides the distillation mechanism is further introduced.
	
	\vspace{-0.3cm}
	\subsection{Semantic Confidence Mechanism}

	Human beings can hardly understand the semantics of different sentences at only one glance. This is because the semantic structures 
	can be different even when these messages indeed tell the same thing. To express a concept also deserves multiple times of processing and correcting. A semantic model, by analogy, also faces the same challenge. Therefore, to model the continuity of semantics expressing and understanding, we enable a model to evaluate its semantic embeddings in both encoder and decoder by the means of semantic confidence. When the semantic confidence reaches a pre-defined threshold, the model releases the processed information for a further down-stream processing. Otherwise, the semantic confidence module sends the embeddings back to the distillation part and asks for a refinement. 
	
	\vspace{-0.5cm}

	\vspace{-0.1cm}
	\subsection{Semantic Distillation Mechanism}
	The distillation mechanism allows a message to be encoded and decoded multiple times under a from-coarse-to-fine structure. The detailed distillation steps are illustrated in the purple and blue blocks in \Figure \ref{Fig:JSNC}. In a practical system, the semantic confidence module and basic encoder/decoder can usually be implemented as artificial neural networks. Once a given message is projected into the feature space (shown in the orange part in \Figure \ref{Fig:JSNC}), the distillation process starts until reaching the maximum distillation times $N$ (representing the cognition ability of a given system) or when the system believes that the semantic extraction is ready for a further processing. 
	In this way, we encourage the model to generate a higher confidence value as mentioned in the semantic confidence mechanism that indicates whether the encoding or decoding procedure is sufficient. The proposed semantic distillation mechanism automatically adjusts the distillation times so as to deal with sentences with both ``hard'' and ``easy'' semantics, and reacts to the varying channel state in an adaptive way. 

	\vspace{-0.3cm}
	\subsection{Implementation Cost and Computational Complexity}
	One major concern of the proposed JSNC approach comes from the computational and implementation cost. We here take the commonly-studied sentence transmission task as an example. Denoted by $L$ the sentence length, and $D$ the embedding dimension, the computational complexity for semantic confidence mechanism is $O(LD)$, which is almost negligible when compared with that of a basic encoder such LSTM ($O(LD^2)$) and Transformer ($O(L^2D)$). The major computations come from the  $N$-times semantic distillation process. It adds the complexity with a linear factor $N$, same as existing recurrent transmission schemes like HARQ \cite{jiang2021deep}. The extra computational cost can be viewed as a case where more computations are adopted to save the transmission time and to pursue a higher accuracy. However, compared with HARQ schemes which have the $O(N)$ time complexity ($N$-times transmission over a channel), JSNC will always take the $O(1)$ complexity since the distillations all happen at a local machine. For implementation purpose, the proposed approach is also backward compatible with existing infrastructures and does not require an elaborately designed protocol.\footnote{Multiple times of semantic distillation does bring certain computational burden. However, it is also notable that neural network-based applications are now benefiting from advanced AI chips, and could be further simplified with existing techniques like neural network pruning, quantization and knowledge distillation.}
	
	\vspace{-0.1cm}
	\begin{figure*}[b]
		\centering
		\setlength{\abovecaptionskip}{-0.1cm}
		\setlength{\belowcaptionskip}{-0.1cm}
		\includegraphics[width=18.1cm]{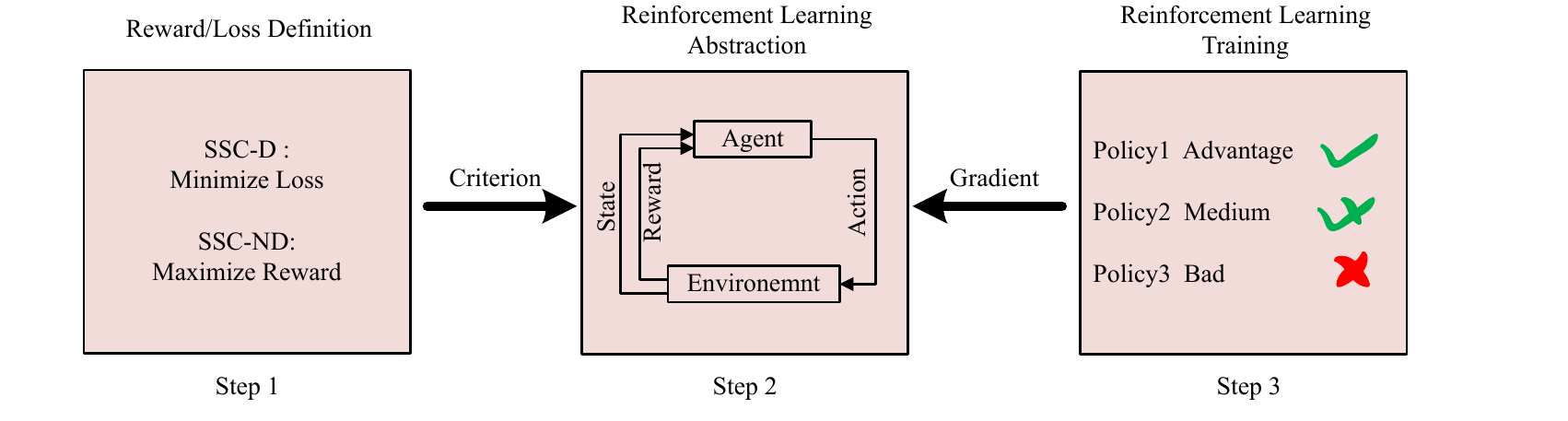}
		\caption{Architecture of the proposed similarity-targeted semantics-persevering communication (SSC) mechanism. By setting the reward function as the expected semantic similarity, a reinforcement learning (RL)-based training scheme is further developed to provide gradients for the established learning system. These gradients are targeted to steer the whole system into a policy improvement, which ultimately lead to a semantic communication system customized by the user.}
		\label{Fig:SSC}
	\end{figure*}
	
	\section{Semantic Communication: Similarity-targeted Semantics-persevering Communication (SSC)}
	Distinguished from semantic coding, a semantic communication system aims to convey the semantic meanings. 
	Consequently, the commonly used CE loss or MSE is flawed as it treats each bit (or word/pixel) with the same weight, which is obviously inconsistent with the nature of languages or other kind of messages. To enable a semantics-level transmission, we put forward a new learning scheme for semantic communication that minimizes the semantic distance, or maximizes the semantic similarity of the sent and received messages.
	
	The overview of our SSC model is illustrated in \Figure \ref{Fig:SSC}. Here we take into account both the differentiable and non-differentiable situations.
	
	\vspace{-0.3cm}
	\subsection{Semantic Communication with Differentiable Semantic Similarity (SSC-D)}
	When a semantic similarity metric is differentiable w.r.t. the message, the gradient will back-propagate through the whole transceiver without any difficulty. In this case, we can simply define the semantic distance as one negative to the similarity metric. For an applicable communication system, this problem then requires a task-specific semantic similarity metric, so long as it is differentiable. For example, we can only transmit the semantically salient parts or the region-of-interest parts using the off-the-shelf computer vision or natural language processing tools, where a masked MSE or CE loss can serve as the semantic distance like what we usually do in semantic compression/segmentation. Since the specific similarity measurement is heavily reliant on the given task, and the gradient back-propagation process will not face any difficulty, we will not give a detailed discussion here. To the best of our knowledge, existing works on semantic communication mostly fall into this category. Instead, we are more interested in the challenging non-differentiable situation, which has not been typically investigated in existing JSCC works. 
	
	\vspace{-0.2cm}
	\subsection{Semantic Communication with Non-differentiable Semantic Similarity (SSC-ND)}
	In real-life scenarios, however, not all semantic metrics are differentiable, and should be treated with a suitable design. Here again we take the sentence transmission task as an example. The widely-used NLP metric like BLEU (Bilingual Evaluation Understudy, which is also used to evaluate a semantic coding system in existing works as in \cite{xie2021deep, jiang2021deep}) is based on the statistical structure information of two paired sentences, and is not differentiable. The commonly used gradient descent method is intractable to implement as a result, and so are the existing deep learning-based JSCC systems like \cite{o2017introduction, bourtsoulatze2019deep, xie2021deep}. In fact, non-differentiable cases may even be more frequently seen, and make this problem more urgent and important.

	To address this non-differentiable problem, a direct and naive solution would leverage surrogate loss functions. However, one may need to redesign a new surrogate module every time the environment or similarity metric changes, and the concrete implementation also requires much expert knowledge. Although substituting the non-differentiable part is promising to provide a quick and light-weight solution, the surrogate process is not universal and lacks scalability in real applications. Instead, we establish a reinforcement learning (RL)-based solution as depicted in \Figure \ref{Fig:SSC}, which provides a model-agnostic, target-agnostic and therefore a universal solution for both the SSC-D and SSC-ND problems.\footnote{RL is one of the possible approaches, and one may also investigate some other solutions. It is also plausible to use surrogate ones if tested promising. Using RL may introduce minor changes to the original model architecture (see SSC-ND in the image transmission task below).} As a well-known technique, reinforcement learning leverages the reward function to encourage an agent to take optimal actions in an interactive environment. In this situation, the reward function is defined as any one that takes a scalar value as the output, thus relaxing the need for its differentiability. The goal of reinforcement learning is to maximize the cumulative reward, which in our case is the semantic similarity. Equipped with RL, we now turn the nontrivial problem into two more friendly sub-problems - how to define an RL process for the semantic encoding-decoding process (more precisely the decoding process as it decides which semantic token to generate), and how to develop a specific learning algorithm. The whole processes transforming a common semantic communication problem into an RL-based one can be summarized in the following steps, as illustrated in \Figure \ref{Fig:SSC}:
	
	\textbf{Step 1}: define a proper semantic similarity score.
	 
	\textbf{Step 2}: transform the learning problem into a reinforcement learning problem.
	 
	\textbf{Step 3}: optimize the accumulated reward function in an RL-based way.
	
	Similar to SSC-D, defining a proper semantic similarity follows the common procedure; thus we will not detail it here. For Step 2, we consider the sentence and image transmission tasks, as they are the most commonly seen scenarios. To define the \textit{state} in an RL-based process, the decoding process needs to be first transformed into a recurrent process. This is easy for a sentence as we can directly use the hidden state of decoder and the previously generated words as the \textit{state}. But for images, the original pixel-level regression pipeline no longer holds under an RL setting. Here we exemplify a pixel-level recurrent decoding scheme by increasing or decreasing the pixel value with a small number so that the \textit{state} can be defined as the intermediate decoded image. Transforming a regression-based decoding scheme into an RL-based one is also feasible to implement in practice but only requires more decoding times. We need to note that although this RL procedure is designed for non-differentiable reward functions, it is also applicable to integrate with differentiable ones, and makes it a general-purpose solution for semantic communication.
	For Step 3, we argue that plenty of existing RL algorithms can be used. As in the case of both the sentence and image transmission tasks for example, an actor-critic RL method is found to be effective and simple enough for implementation. 
	
	\vspace{-0.5cm}
	\subsection{The Information Overhead}
	In this subsection, we analyze the information overhead imposed to the proposed semantic communication system. Considering a specific communication scenario, we can divide the information overhead into three parts:
	\begin{itemize}
	\item{\textit{Semantics part.}} It is the amount of information necessarily required to convey the semantic meanings.
	\item{\textit{Recurrence part.}} On top of the semantics part, the extra information needed to enable a precise recovery (that enables a perfect bit-level accuracy).
	\item{\textit{System part.}} Some content-agnostic parts which are regarded as the system burden.
	\end{itemize}

	The most significant feature that characterizes semantic communication lies in that only the semantics part is necessarily needed to transmit, which indicates semantic communication is beyond Shannon's reliable communication \cite{strinati20216g} and theoretically brings a lower information overhead. It is noteworthy that the proposed RL-based SSC system provides the first universal and practical solution towards this goal, and can therefore achieve superior performance if under the same overall information overhead. In contrast, the conventional reliable communication can be viewed as a special case of semantic communication, where all the bits are recognized as necessary without a semantics-particular consideration.
	
	In a practical semantic communication system, the process of semantics extraction also consumes certain computational resources and brings extra informational computations locally. As we will show in the experimental results, the proposed SSC scheme better conveys the semantic meanings when compared with the conventional model under the same information and computation overhead.

	\vspace{-0.1cm}
	\section{Case Study}	

	\vspace{0.cm}
	\begin{figure*}[b]
		\centering
		\setlength{\abovecaptionskip}{0.cm}
		\setlength{\belowcaptionskip}{-0.cm}
		\includegraphics[width=18.1cm]{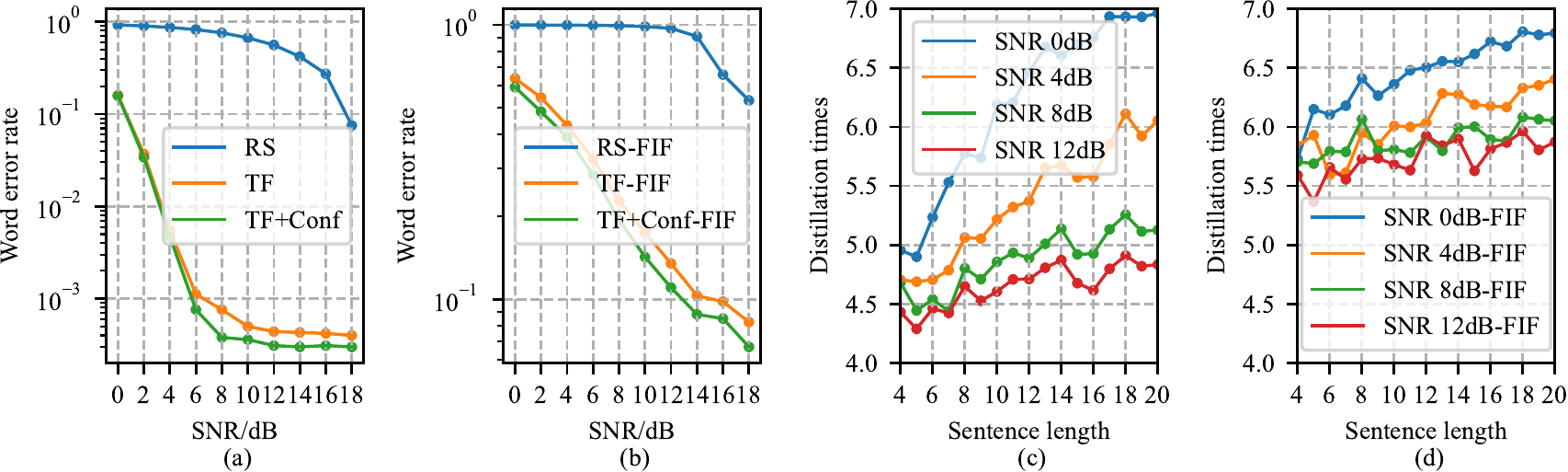}
		\caption{Numerical results of the proposed confidence-based joint semantics-noise coding mechanism. (a) and (b) exhibit the word error rate versus the examined SNR level in AWGN channel and phase-invariant fading (FIF) channel respectively. (c) and (d) show the average distillation times when tested under different input sentences and channel states.}
		\label{Fig:NUM_SCoding}
	\end{figure*}
	
	\subsection{Datasets}
	We use European-parliament dataset for sentence transmission, and CIFAR-10 for image transmission. Apart from these existing datasets, we also propose a new real communication scenario to examine the performance of semantics-aware communication, which we call it ``speak something''. 
	We carry numerical experiments on both the AWGN and phase-invariant fading (denoted as FIF in this article) channels (for demonstration purpose, we do not use any channel adaptation technique, though we note some other channels and adaptation techniques are also plausible and  better complement our solutions) to simulate the real wireless communication scenarios.
	
	\vspace{-0.3cm}		
	\begin{figure*}[t]
		\centering
		\setlength{\abovecaptionskip}{0.cm}
		\setlength{\belowcaptionskip}{-0.cm}
		\includegraphics[width=18.1cm]{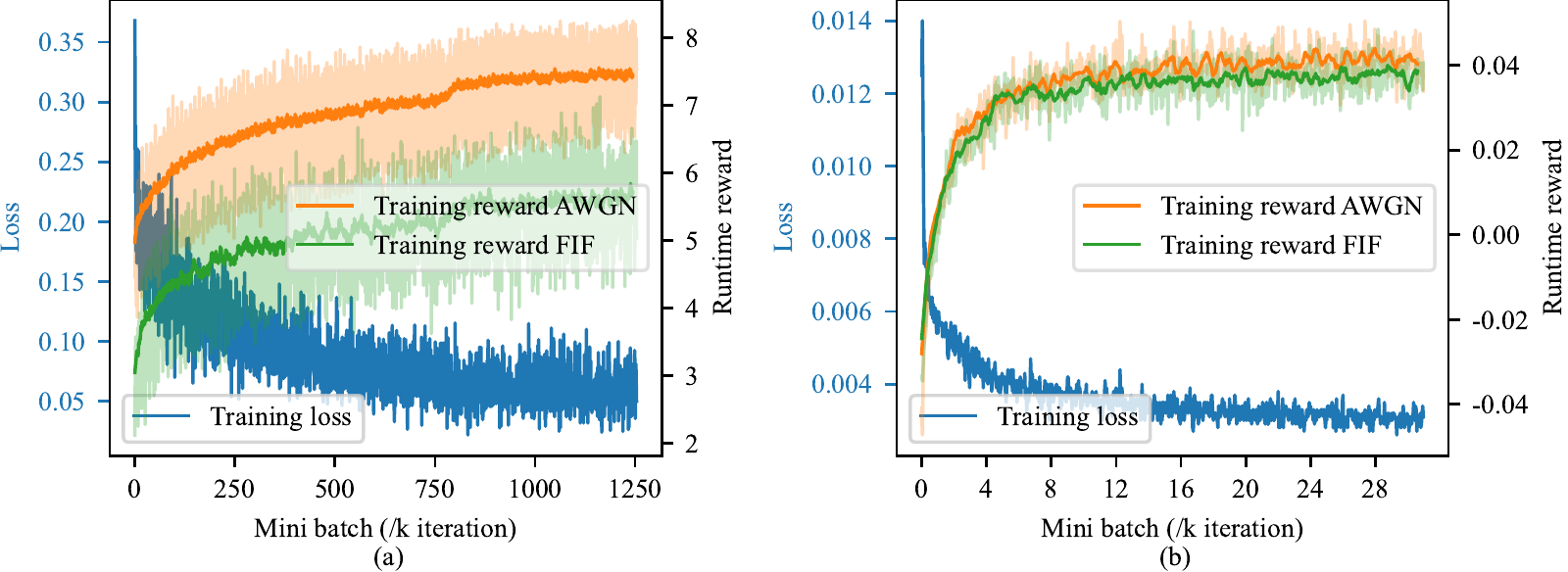}
		\caption{Learning curves of the proposed RL-based semantic communication system (SSC-ND). (a) Training curves on the sentence transmission (European-parliament) task. (b) Training curves on the image transmission (CIFAR-10) task. Runtime reward curves in AWGN and phase-invariant fading (FIF) channel are shown in orange and green lines respectively. The left y-axis measures the loss, while the right one measures the runtime reward.}
		\label{Fig:NUM_SComm}
	\end{figure*}

	\subsection{Semantic Coding}
	We focus on the commonly-used European-parliament dataset for the semantic coding task. A Transformer structure is adopted as the basic encoder since it has shown great performance in existing works \cite{xie2021deep, jiang2021deep}. 
	First we provide a holistic view on the proposed confidence-based distillation technique in \Figure \ref{Fig:NUM_SCoding}(a-b), where one can obviously find that the proposed joint semantics-noise coding method shows its superiority over the rigid one-time encoding-decoding schemes consistently, and further reduces the bits required for an acceptable transmission. Under a 10dB AWGN channel for example, the proposed distillation process (\ie, TF+Conf) achieves a word error rate of 0.038\%, which is 24\% lower than the Transformer baseline (\ie, TF), and significantly lower than the conventional separate encoding-decoding method like Reed-Solomon (RS) code of 5-bit fixed length. When tested on the phase-invariant fading channel as illustrated in \Figure \ref{Fig:NUM_SCoding}(b), the model is still robust when compared with that without a distillation mechanism.
	
	Aside from the numerical simulation results, we provide a detailed insight into the semantics-noise-robust coding mechanism, showing how the model interprets the semantics and why it is semantics-aware. 
	First, we find that the model learns to pay more attention to those with more complex semantic structures, where the distillation time positively changes as the sentence length increases, as shown in \Figure \ref{Fig:NUM_SCoding}(c-d). Moreover, when the channel state degrades in terms of SNR (signal-to-noise ratio) level, the proposed JSNC mechanism succeeds in performing more distillations to adaptively explore a more in-depth semantic representation, which is similar to what we human beings do. Typically we find that an FIF channel generally leads to higher distillation times than an AWGN one, 
	which also suggests that the proposed JSNC approach better captures the semantics under a more complex channel state. Also, we note that under extreme cases where the useful information is submerged in large random noise (as shown in \Figure \ref{Fig:NUM_SCoding}(d)), the distillation times tend to saturate since learning the semantics gets much more difficult.

	\vspace{-0.3cm}
	\begin{table*}[t]
		\centering
		\caption{Real-life examples on the proposed similarity-targeted semantics-persevering communication (SSC) mechanism.}
		\label{tab:SSC}
		
		\def\arraystretch{1.0} 
		
		\begin{tabular}{cc}	
			\hline
			
			\multirow{5}{*}{(a)}&\multicolumn{1}{|l}{INPUT: \textit{to further understand and certify the effectiveness of the proposed semantic communication system}}\\
			\multicolumn{1}{c}{~}&\multicolumn{1}{|l}{\text{CE1: to further this and audits the effectiveness of the sometimes detention communication system}}\\
			\multicolumn{1}{c}{~}&\multicolumn{1}{|l}{\text{CE2: to further this and phase the effectiveness of the proposed prosecution communication fraude}}\\
			\multicolumn{1}{c}{~}&\multicolumn{1}{|l}{\text{RL1: to further understand and that the effectiveness of the proposed that communication system}}\\
			\multicolumn{1}{c}{~}&\multicolumn{1}{|l}{\text{RL2: to further understand and that the effectiveness of the proposed that communication system}}\\
			\hline

			\multirow{5}{*}{(b)}&\multicolumn{1}{|l}{INPUT: \textit{he claimed that he saw a strange building on the street yesterday}}\\
			\multicolumn{1}{c}{~}&\multicolumn{1}{|l}{\text{CE1: he claimed that he saw a strange building on the street yesterday}}\\
			\multicolumn{1}{c}{~}&\multicolumn{1}{|l}{\text{CE2: he claimed that he saw a strange building on the street yesterday}}\\
			\multicolumn{1}{c}{~}&\multicolumn{1}{|l}{\text{RL1: he claimed that he saw a strange building on the street yesterday}}\\
			\multicolumn{1}{c}{~}&\multicolumn{1}{|l}{\text{RL2: he says that he saw a strange building on the street yesterday}}\\
			\hline
			
			\multirow{5}{*}{(c)}&\multicolumn{1}{|l}{INPUT: \textit{he claimed that he saw a strange building walking on the street yesterday}}\\
			\multicolumn{1}{c}{~}&\multicolumn{1}{|l}{\text{CE1: he claimed that he saw a strange rather curious on the street yesterday}}\\
			\multicolumn{1}{c}{~}&\multicolumn{1}{|l}{\text{CE2: he claimed that he saw a strange building site on the street yesterday}}\\
			\multicolumn{1}{c}{~}&\multicolumn{1}{|l}{\text{RL1: he claimed that he saw a strange building that on the street yesterday}}\\
			\multicolumn{1}{c}{~}&\multicolumn{1}{|l}{\text{RL2: he reminded that he saw a strange building that on the street yesterday}}\\

			\hline
			
		\end{tabular}
		\vspace{-0.3cm}
	\end{table*}

	\subsection{Semantic Communication}
	We exemplify two widely-investigated scenarios, \ie, the sentence and image transmission tasks (shown in \Figure \ref{Fig:NUM_SComm}(a) and (b) respectively) to illustrate the feasibility of developing such a semantic communication system, regardless of the differentiability of semantic similarity. We set the reward function for the sentence transmission task as the CIDEr-D (CIDEr: Consensus-based Image Description Evaluation) score and MSE gain (the present MSE score versus the MSE value in the next time step, so that a positive MSE gain means a quality improvement) for the image paradigm. 
	An actor-critic algorithm is used to provide the gradient for the whole system in a self-supervised way. In \Figure \ref{Fig:NUM_SComm}, the training reward under a 10dB AWGN channel is reported in the orange lines, and that under FIF channel is given in green lines. We only plot the training loss under AWGN channel for figure tidiness.

	As we can observe from the learning curves in \Figure \ref{Fig:NUM_SComm}, the proposed SSC-ND approach jumps out of the initial local minima and converges stably as training proceeds, which convincingly illustrates the feasibility of this innovation. With SSC-ND, we enable the communication system to develop a semantics-level ability and dive into a deeper level of communication. It minimizes the semantic difference of any two messages, instead of securing a bit-level accuracy.
	We note that training under the phase-invariant fading channel brings more challenges for the learning system, but the proposed approach can still behave properly. These observations point out that the proposed semantic communication approach is also robust and promising if deployed in real scenarios.
	
	To further understand and certify the effectiveness of the proposed semantic communication system, we provide some examples on the sentence-based model in Table \ref{tab:SSC}. The same neural network trained on conventional cross entropy loss is adopted here as the baseline, and we feed the same sentence multiple times to diminish the impact of channel noise.\footnote{Due to the page limit, we give two examples each.}
	We find that by minimizing the semantics-level distance, the model develops an interesting pattern that differs significantly from the conventional model. It favors words and phrases with similar meanings at the expense of certain degradation in bit-level accuracy, and favors sentences more likely to happen at the semantic level. In Table \ref{tab:SSC} (a) for example, the proposed model better captures the key meanings like ``understand'' and ``communication system'', while the baseline model distorts what we want to express severely. We also find that our model can properly choose a similar word even an error does occur, like substituting ``claimed'' into ``says'' in Table \ref{tab:SSC} (b). Even under extreme conditions where a semantically insane sentence is exposed the system, SSC-ND can still exhibit a robust performance and better catches the key idea, as illustrated in the last example of Table \ref{tab:SSC}.

	\subsection{Real Communication Scenario: Speak Something}
	Aside from the above experimental results, we further introduce a new communication scenario in this article to examine both the performance and effectiveness of the proposed approaches. Our inspiration comes from a commonly played game, ``draw something'' (or ``speak something''), which requires a speaker to depict a given object to a remote listener, but does not allow the speaker to directly tell what it is. In our experiment, the listener is allowed to make three different final assumptions, which are sorted in descending order of confidence. Correspondingly, we score these possible assumptions with bonus of 3, 2, and 1 respectively if the corresponding one proves to be correct. As we find in the experiment, the proposed SSC scheme yields a score of around 2.3, which is significantly higher than the conventional baseline whose score is around 1.2. Similarly, when we ask the speaker to add one cue for each turn, and compare the number of turns required to correctly recognize the given object, a fewer transmission turns are also observed. The improvement brought by JSNC approach is not that significant in this experiment, since the absolute accuracy boost is still limited and saturates when close to 1.

	\section{Open Challenges with the Proposed Approach}

	\textbf{Theory on Semantic Information.} A critical cornerstone for semantic communication lies in semantic information theory. Besides serving as the concrete measurement of semantic information bound (\ie, the semantics part in the information overhead) and an important metric for evaluating the semantic abstraction ability, semantic information also provides a strong guidance for the proposed SSC scheme. Though researches on practical semantic communications have seen much growth on simple semantic concepts, theorizing a universal semantic information theory still has a long way to go.
	
	\textbf{Semantics and Security Design.}
	A semantics-aware communication system, in spite of its promising capability in revealing the underlying meanings, also raises some new concerns from the security perspective. The abstraction nature of semantics and target-oriented property (for example, the same semantic token can bring different meanings for a group of receivers and for different tasks) enable a highly-encrypted, receiver-selective and flexible enough solution for next generation communication, which can further get integrated with existing physical-layer encrypting techniques \cite{yang2015safeguarding}. Although researches on the semantic security are limited in the literature, we highlight the potential a semantics-aware communication brings, and also point out some possible challenges like the user preference, the implementation cost \etc.
	
	\textbf{Challenges from the Implementation Perspective.}
	As categorized into computation-based methods, the proposed scheme also faces certain implementation challenges, such as the computation-performance trade-off, demanding from AI-based infrastructures and task-specific generalizations. However with the hardware support, we believe that semantics-aware communications provide an important opportunity for beyond 5G and 6G applications.
	
	\section{Conclusion and Outlook}
	The core contributions of this article can be summarized into three key points. First, we provide a holistic overview on the existing semantics-aware communication techniques, and analyze the underlying drawbacks of the current schemes. Second, aiming at solving the aforementioned problems, we devote to rethink what a semantics-aware communication system ought to be, and detail some related concerns like implementation cost and future designs. Third, we establish our joint semantics-noise coding (JSNC) solution for the semantic coding problem, and an RL-based similarity-targeted semantic communication mechanism for both differentiable and non-differentiable semantic similarity metrics (\ie,  SSC-D \& SSC-ND), to demonstrate the feasibility of building such a semantics-aware communication system and its promising benefits. 
	
	We notice that almost all the existing semantics-aware works fall into the semantic coding part, but only a few concentrate on the emerging topic of semantic communication, and are restricted on the differentiable level. We hope this work could provide a new insight for future semantics-aware communication systems. 

	\bibliographystyle{IEEEtran}
	\bibliography{Magazine_KunLu}

	
	
	
	

\end{document}